\documentclass[preprint]{aastex631}
\usepackage{graphicx}
\usepackage{natbib}
\usepackage{amsmath}
\usepackage{here}

\accepted{July 14th, 2025}

\submitjournal{ApJ}

\begin{document}

\title{Stellar Models Also Limit Exoplanet Atmosphere Studies in Emission}

\author[0000-0002-5967-9631]{Thomas J. Fauchez}
\affiliation{NASA Goddard Space Flight Center, 8800 Greenbelt Road, Greenbelt, MD 20771, USA}
\affiliation{Integrated Space Science and Technology Institute, Department of Physics, American University, Washington DC}
\affiliation{Sellers Exoplanet Environment Collaboration (SEEC), NASA Goddard Space Flight Center}
\affil{Consortium on Habitability and Atmospheres of M-dwarf Planets (CHAMPs), Laurel, MD, USA}
\author[0000-0002-7008-6888]{Elsa Ducrot}
\affiliation{LESIA, Observatoire de Paris, CNRS, Universit\'e Paris Diderot, Universit\'e Pierre et Marie Curie, 5 place Jules Janssen, 92190 Meudon, France}
\affiliation{AIM, CEA, CNRS, Universit\'e Paris-Saclay, Universit\'e de Paris, F-91191 Gif-sur-Yvette, France}
\author[0000-0002-3627-1676]{Benjamin V. Rackham}
\affiliation{Department of Earth, Atmospheric and Planetary Sciences, Massachusetts Institute of Technology, Cambridge, MA 02139, USA}
\affiliation{Kavli Institute for Astrophysics and Space Research, Massachusetts Institute of Technology, Cambridge, MA 02139, USA}
\author[0000-0002-7352-7941]{Kevin B. Stevenson}
\affiliation{Johns Hopkins Applied Physics Laboratory, Laurel, Maryland 20723, USA}
\affil{Consortium on Habitability and Atmospheres of M-dwarf Planets (CHAMPs), Laurel, MD, USA}
\author[0000-0002-4321-4581]{L. C. Mayorga}
\affiliation{Johns Hopkins Applied Physics Laboratory, Laurel, Maryland 20723, USA}
\affil{Consortium on Habitability and Atmospheres of M-dwarf Planets (CHAMPs), Laurel, MD, USA}

\author[0000-0003-2415-2191]{Julien de Wit}
\affiliation{Department of Earth, Atmospheric and Planetary Sciences, Massachusetts Institute of Technology, Cambridge, MA 02139, USA}
\affiliation{Kavli Institute for Astrophysics and Space Research, Massachusetts Institute of Technology, Cambridge, MA 02139, USA}

\correspondingauthor{Thomas J. Fauchez}
\email{thomas.j.fauchez@nasa.gov}

\begin{abstract}

Stellar contamination has long been recognized as a major bottleneck in transmission spectroscopy, limiting our ability to accurately characterize exoplanet atmospheres---particularly for terrestrial worlds. In response, significant observational efforts have shifted toward emission spectroscopy as a potentially more robust alternative, exemplified by initiatives such as the 500-hour JWST Rocky Worlds Director’s Discretionary Time (DDT) program. However, the extent to which emission spectroscopy may be affected by stellar effects remains mostly unexplored, in stark contrast with the extensive exploration and mitigation work for transmission spectroscopy.

In this study, we assess the impact of imperfect knowledge of stellar spectra on exoplanet atmospheric retrievals from emission spectroscopy. At 12.8\,$\micron$, none of the considered bare surface types---basalt, ultramafic, Fe-oxidized, and granitoid---can be reliably distinguished when accounting for the 3$\sigma$ model precision between SPHINX and PHOENIX. At 15.0\,$\micron$, only the granitoid surface is distinguishable from all others above this threshold. These results show that stellar model uncertainty alone substantially limits our ability to constrain surface composition from photometric data, even before including other sources of uncertainty such as stellar radius.

Also, we find that current 15\,$\micron$ eclipse depth estimations using different stellar models introduce a 60\,ppm difference for M8 and 20\,ppm for M5 stars. This model discrepancy leads to a degeneracy in retrieved planetary albedos and weakens constraints on the presence of an atmosphere. We therefore recommend that future JWST secondary eclipse observations systematically include stellar mid-infrared spectroscopy to mitigate these uncertainties.

\end{abstract}


\section{Introduction} \label{sec:intro}

\begin{figure*}[t]
    \centering
    \includegraphics[width=0.9\textwidth]{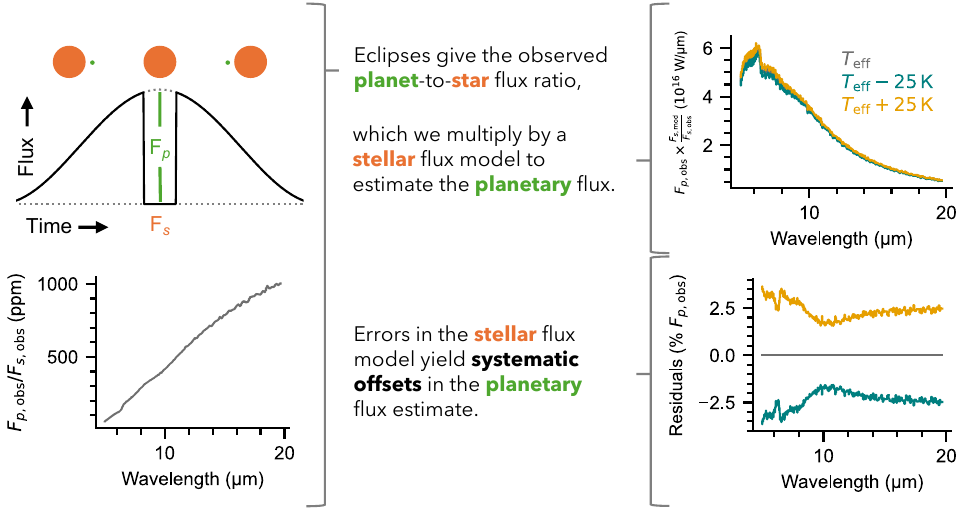}
    \caption{
        Schematic of an exoplanetary secondary eclipse and associated measurements.
        The principal observable of a secondary eclipse is the planet-to-star flux ratio in a given bandpass (left).
        This must be multiplied by a model of the stellar flux to estimate the planetary flux (top right).
        Any errors in the stellar flux model, due to limitations in model precision, errors in the effective temperature estimate, or unaccounted photospheric inhomogeneities (e.g., spots and faculae), will effectively bias the calculated planetary flux (bottom right).
    }
    \label{fig:schematic}
\end{figure*}

The interpretation of exoplanet transmission spectra is fundamentally limited by our understanding of the stellar flux illuminating the planetary atmosphere.
In particular, the inhomogeneity of a stellar photosphere imprints spectral features on measured transmission spectra, a phenomenon termed the ``transit light source (TLS) effect'' \citep{Rackham2018, Rackham2019}.
This effect arises because, due to photospheric heterogeneities such as spots and faculae, the stellar flux illuminating the planetary atmosphere may differ from the disk-integrated stellar spectrum used to normalize transit depths.
Transit observations over the past two decades have shown these stellar spectral imprints \citep{Pont2008, Sing2011, McCullough2014, Rackham2017, Espinoza2019, Kirk2021}, with recent JWST observations providing the clearest evidence \citep{Lim_2023, May_2023, Moran2023, Cadieux_2024, Radica2025, Rathcke2025}.
It is now widely understood to be a major limiting factor for precise atmospheric characterization \citep{Iyer2020, Rackham2023, Rackham2024}, particularly for small exoplanets, which present the most minute atmospheric features. Insights from years of theoretical and observational explorations of the TLS effect, combined with the development of a series of mitigation strategies \citep[e.g.,][]{Wakeford2019,Garcia2022,Berardo2024, Narrett2024, Rathcke2025}, have recently coalesced into a roadmap for the robust atmospheric study of terrestrial worlds \citep{TJCI2024}. 

As emission measurements are thought to be unaffected by the TLS effect, they have been presented as an alternative to transmission measurements for revealing the presence of atmospheres around terrestrial worlds via measurements of a cooler planetary dayside \citep[due to heat redistribution to zeroth order;][]{Koll_2022}. Based on this understanding, the community supported the request for the Rocky Worlds 500-hr DDT program \citep{Redfield2024} that will collect mid-infrared observations of secondary eclipses of up to a dozen transiting exoplanets.  However, while a good deal of recent work has centered on overcoming this challenge for exoplanet transmission measurements, much less attention has been paid to how our incomplete knowledge of the stellar flux affects exoplanet emission measurements \citep{ducrot_2024, Coy_2024}.

Emission spectra are obtained by comparing the combined light of the star and planet just before secondary eclipse to the light of the star alone during eclipse \citep{Seager1998, Marley1999, Sudarsky2000}.
As shown in \autoref{fig:schematic}, the primary observable is the planet-to-star flux ratio ($F_\mathrm{p, obs} / F_\mathrm{s, obs}$), which must be multiplied by a value for the measured stellar flux inside the eclipse  ($F_\mathrm{s, obs}$) to estimate the absolute planetary flux  ($F_\mathrm{p, obs}$). In practice, no stellar model is required to that point as done in \cite{Zieba_2023,Greene_2023,ducrot_2024}. However, to derive planetary properties from $F_\mathrm{p, obs}$ or $F_\mathrm{p, obs} / F_\mathrm{s, obs}$, a stellar spectrum is required to generate planetary emission models (bare surfaces and atmospheres)  as shown in \cite{ducrot_2024} (their Figure 2). Note that some studies directly multiplies the planet-to-star flux ratio by a stellar model (PHOENIX/BT-Settl in \cite{Coulombe_2023} and SPHINX in \cite{August_2024}) to estimate the stellar flux inside the eclipse and therefore to retrieve $F_\mathrm{p, obs}$. In any case, discrepancy between the true stellar emission spectrum and the assumed stellar flux---whether due to inherent model limitations or unmodeled photospheric heterogeneity---will propagate into systematic errors in the inferred planetary emission spectrum (\autoref{fig:schematic}).
This issue is now especially relevant as the community embarks on detailed emission studies of small exoplanets, including the Rocky Worlds DDT program, which aims to characterize the thermal emission of transiting terrestrial exoplanets.
The success of such programs depends critically on the accuracy of stellar models used to interpret planetary flux measurements.

Here we examine the impact of this limitation on planetary emission measurements.
We focus on the example of TRAPPIST-1 (2MASS J23062928-0502285), an M8 dwarf hosting seven transiting terrestrial exoplanets \citep{Gillon2016, Gillon2017, Luger2017} that are similar in size, mass, and irradiation to the Solar System planets \citep{Agol2021}.
Providing an ideal laboratory to study how the atmospheric evolution of multiple planets orbiting an M dwarf can impact their habitability \citep{Lincowski2018, Krissansen-Totton_2022}, this system has been the focus of extensive JWST observations in both transmission and emission \citep{Greene_2023, Lim_2023, Zieba_2023, ducrot_2024, Radica2025, Rathcke2025}. The emission observations have  highlighted the significant shortcomings of SPHINX \citep{SPHINX} or PHOENIX \citep{BT-Settl} models (see section \label{sec:bias}) into accurately modeling TRAPPIST-1 mid-IR spectrum \citep{Ih2023,Zieba_2023,ducrot_2024} and, unfortunately, no MIR spectroscopic observation of the star has been acquired prior to the search for secondary atmospheres on its surrounding planets. Note that similar searches have been done for WASP-18\,b \citep{Coulombe_2023} and LHS~1478\,b \citep{August_2024} but, unfortunately, no stellar fluxes were provided in this studies, therefore prohibiting comparisons to synthetic stellar spectra. 

This paper is organized as follows.
We show in Section \ref{sec:bias} that the mid-infrared stellar models currently used in the community inherently lead to biases in exoplanet parameters retrieved from secondary eclipse observations.
We then briefly discuss and provide our conclusions and recommendations for the community in Section \ref{sec:conclusions}.

\section{Quantifying Star-Induced Biases in Exoplanet Emission Studies}\label{sec:bias}

\subsection{Performance Requirement}
\begin{figure}[ht!]
\centering
\includegraphics[width=0.9\columnwidth]{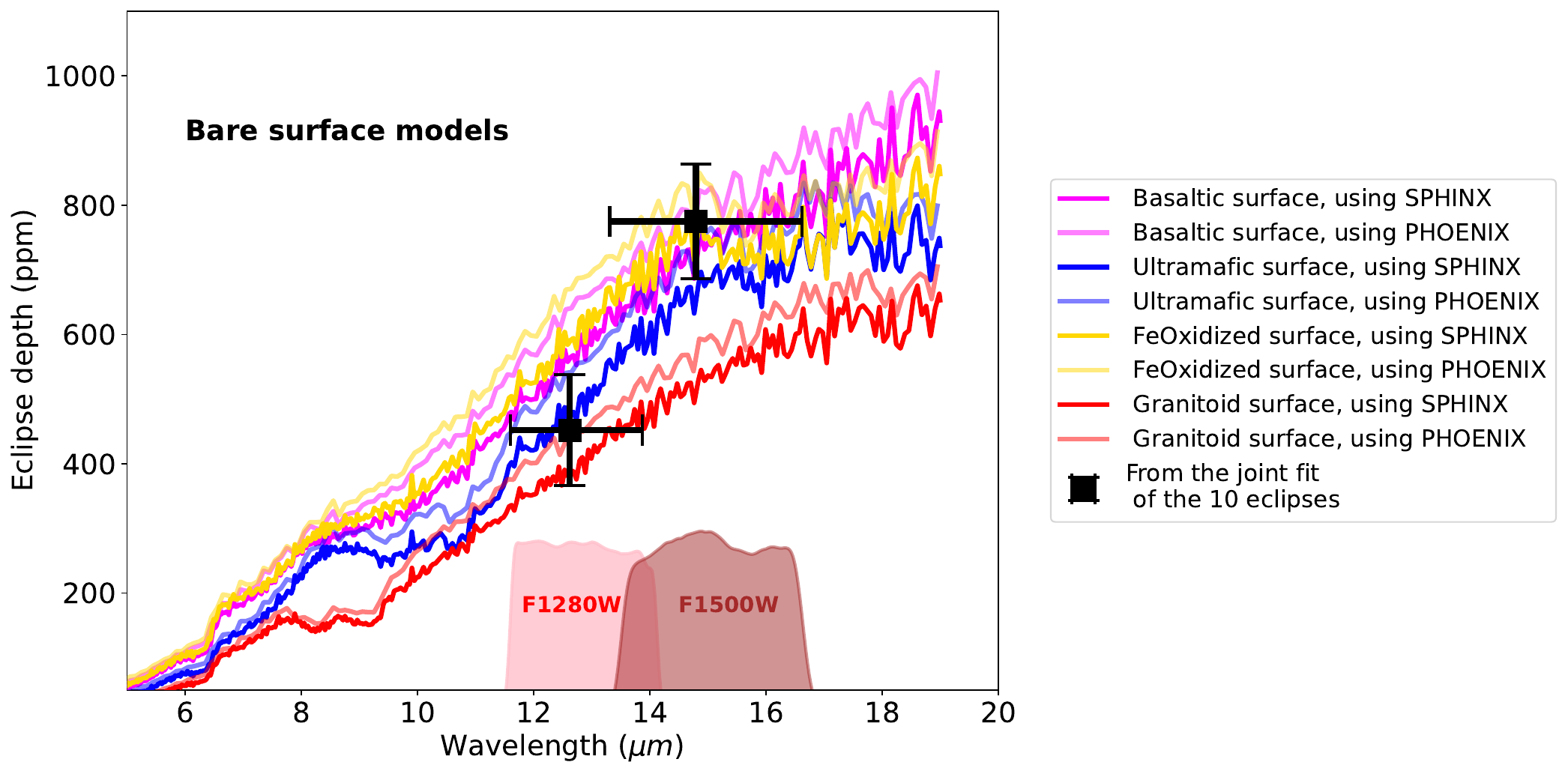}
\includegraphics[width=0.9\columnwidth]{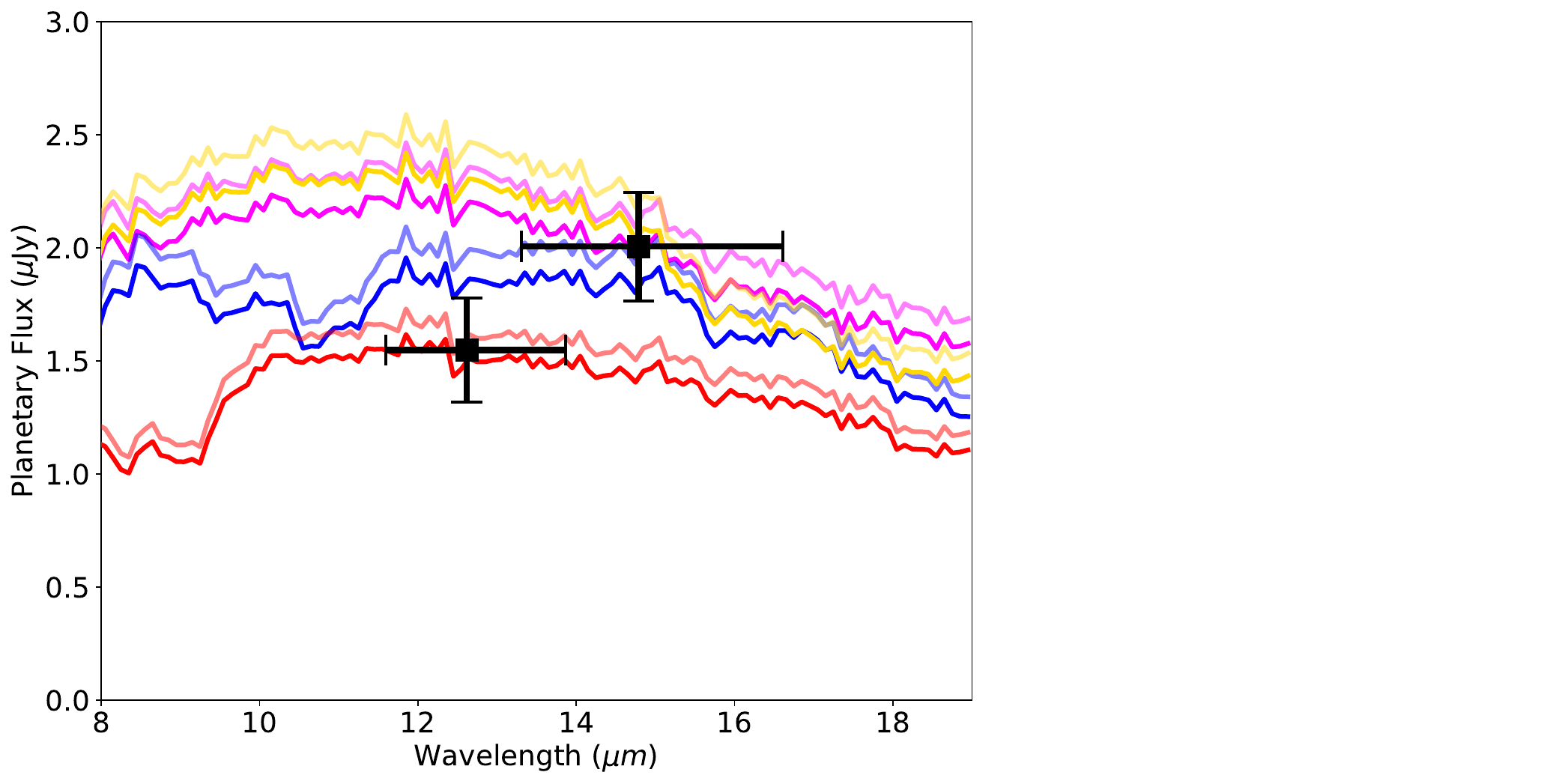}
\caption{\textbf{Top row:} Observed eclipse depth $F_\mathrm{p, obs} / F_\mathrm{s, obs}$ (ppm) for TRAPPIST-1 b, derived from the joint analysis of 10 secondary eclipse observations at 12.8 and 15\,$\micron$, with 1\,$\sigma$ uncertainties, based on the fiducial joint MCMC analysis described in \cite{ducrot_2024}. Spectrally dependent eclipse depths for various bare surface models are shown, computed using both the uncorrected SPHINX and PHOENIX stellar models, displayed in distinct colors. The red and brown shaded regions indicate the transmission profiles of the F1280W and F1500W MIRI filters, respectively. The inevitable use of different stellar spectral models to generate bare surface models blurs the interpretation of the two photometric data points.\\
\textbf{Bottom row:} Observed  planetary flux $F_\mathrm{p, obs}$ ($\mu$Jy) for TRAPPIST-1 b, computed from $(F_\mathrm{p, obs} / F_\mathrm{s, obs}) \times F_\mathrm{p, obs}$, assuming a 3\% absolute MIRI photometric precision, as adopted for TRAPPIST-1 c in \cite{Greene_2023}. Even in absolute flux units, stellar spectra remain necessary to model the surface-dependent planetary emission (e.g., \cite{Ih2023}), and model-dependent discrepancies persist.
}
\label{Fig:eclipse}
\end{figure}

Before quantifying the effect of imperfections in existing stellar models for emission-based exoplanet studies, we briefly review the precision required to reach targeted inferences. In \autoref{Fig:eclipse}, which is adapted from Figure 2 of \cite{ducrot_2024}, the upper panel illustrates that the predicted eclipse depth for each surface type varies depending on whether SPHINX (as in \cite{ducrot_2024}) or PHOENIX (as in \cite{Greene_2023} for TRAPPIST-1 c) is used. This model dependence blurs the interpretation of the two photometric data points. A true, measured stellar spectrum would yield eclipse depths distinct from those predicted by SPHINX or PHOENIX, and would not suffer from such discrepancies.
Similarly, even if one considers only the observed planetary flux $F_\mathrm{p, obs}$ ($\mu$Jy), obtained by multiplying the eclipse depth by the observed stellar flux during eclipse $F_\mathrm{p, obs}$ (as in \cite{Greene_2023}, assuming a 3\% absolute MIRI photometric precision), stellar models are still required to capture the surface-type-dependent spectral features needed to interpret the photometric measurements. In this case as well, a true, measured stellar spectrum would eliminate the discrepancies associated with the use of theoretical stellar models.

\begin{figure}[t!]
\centering 

\includegraphics[width=0.8\linewidth]{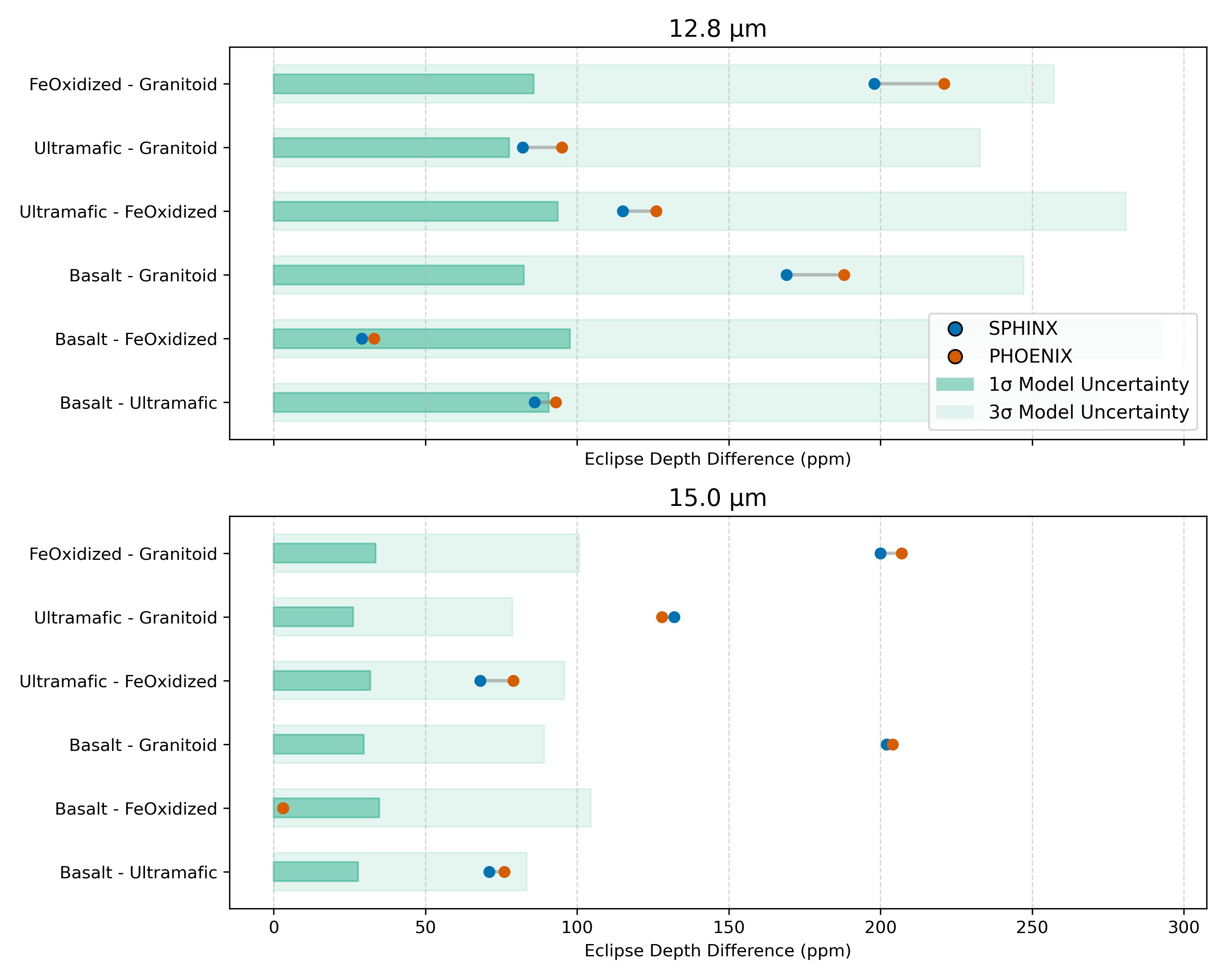}
\caption{Eclipse depth differences to distinguish between bare surface types from \cite{Ih2023} and \cite{ducrot_2024} at 12.8\,$\micron$ (top panel) and 15.0\,$\micron$ (bottom panel) for SPHINX (blue dots) and PHOENIX (orange dots). Each dot represents the predicted eclipse depth difference between two surface types using a single stellar model. The shaded green bands show the uncertainty associated with using different stellar models: the 1\,$\sigma$ (dark green) and 3\,$\sigma$ (light green) levels correspond to the quadrature sum of the model discrepancies for each individual surface in the pair. This illustrates the extent to which stellar model uncertainties alone can limit the ability to distinguish between surface types in photometric secondary eclipse observations, even before considering other sources of uncertainty such as the stellar radius.}
\label{Fig:barchart}
\end{figure}

 \autoref{Fig:barchart} presents the eclipse depth differences between bare surface types from \cite{Ih2023} and \cite{ducrot_2024}, as in \autoref{Fig:eclipse}, for the MIRI 12.8\,$\micron$ (top panel) and 15.0\,$\micron$ (bottom panel) photometric bands, using both SPHINX and PHOENIX. Also shown are the model precisions between SPHINX and PHOENIX at 1\,$\sigma$ (solid green bars) and 3\,$\sigma$ (transparent green bars), computed as the quadrature sum of the model uncertainties for each surface pair.
At 12.8\,$\micron$, none of the surface types are distinguishable at a significance level exceeding the 3\,$\sigma$ stellar model precision. At 15.0\,$\micron$, only the granitoid surface type is distinguishable from all others at a level above the 3\,$\sigma$ threshold. These results demonstrate that stellar model uncertainty alone substantially limits the ability to constrain surface composition from photometric observations, even before accounting for additional sources of uncertainty such as the stellar radius. A true, observed spectrum—assumed to be temporally stable—would eliminate this source of uncertainty.

\subsection{Imperfect Stellar Models}
\subsubsection{Late Type Stars Are Far From Black Bodies}
\begin{figure*}[t!]
\centering 

\includegraphics[width=1.\linewidth]{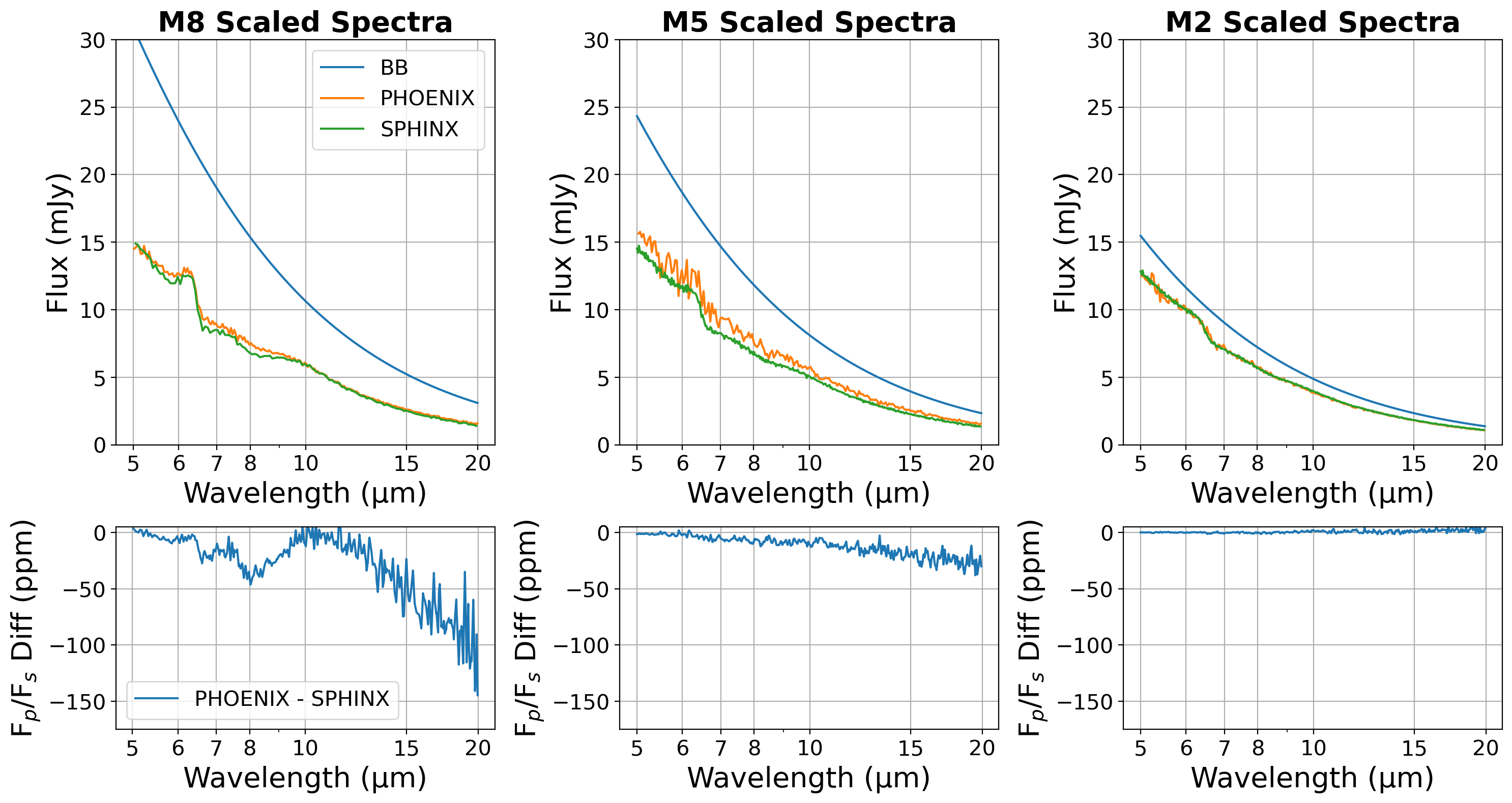}
\caption{Comparisons of stellar model fluxes for M8 (TRAPPIST-1) and a M5 ($T_\mathrm{eff} = 2800$\,K) and M2 ($T_\mathrm{eff} = 3400$\,K) stars and their impact on eclipse depth estimation for a planet like TRAPPIST-1\,b. The various models considered are: blackbody (BB), PHOENIX/BT-Settl \citep{BT-Settl} and SPHINX (\cite{SPHINX}, only available for M dwarfs). \textbf{Top row:} stellar flux (mJy) normalized to the observed 2MASS $J$ = 11.35 mag flux of TRAPPIST-1. 
\textbf{Bottom row:} simulated eclipse depth ($F_\mathrm{p, sim} / F_\mathrm{s, sim}$) difference where $F_\mathrm{p, sim}$ assumes a 490\,K blackbody emission \citep{ducrot_2024} for the planet and the $F_\mathrm{s, sim}$  from each stellar model. For the M8 TRAPPIST-1, the difference reaches 40 and 60~ppm at 12.8 and 15\,$\micron$, respectively.}
\label{Fig:specData}
\end{figure*}

The strongly non-blackbody nature of the spectrum of TRAPPIST-1 in the MIR has been highlighted previously \citep{Ih2023, Zieba_2023,Davoudi_2024}. Using the SPHINX model, these authors estimated $\alpha$, the ratio of the stellar brightness temperature in the MIRI band to the stellar effective temperature. They determined $\alpha = 0.72$ at 14.87\,$\micron$. Additionally, using MIRI photometric observations, \cite{Zieba_2023} measured a stellar flux of 2.599 $\pm$ 0.079\,mJy at 14.87\,$\micron$ and found a very similar value of $\alpha = 0.71\pm 0.02$.  The authors noted that the star's brightness temperature at these wavelengths is $1867 \pm 55$\,K, significantly cooler than its effective temperature \citep[$T_\mathrm{eff}=2569 \pm 28$\,K,][]{Davoudi_2024}. This $\alpha$ value highlights a large departure from a black-body spectrum in the mid-IR wavelength range. For a late-type star, near the brown dwarf transition, such a departure may be due to molecular absorption features, star spots, and/or silicate clouds, as observed for brown dwarfs \citep{Miles_2023}.  
Each of these effects presents a significant challenge for modeling the atmospheres of late-type stars and thus their planets, since the stellar spectrum is an essential input for planetary atmospheric models.

\subsubsection{Late Type Stars Models Differences in the Mid-Infrared}
\autoref{Fig:specData} shows how stellar models differ for an M8 star (like TRAPPIST-1), a M5 ($T_\mathrm{eff} = 2800~K$) and a M2 ($T_\mathrm{eff} = 3400~K$) star. Although differences between the stellar models are noticeable, their departure from the blackbody spectrum in the JWST MIRI (5--28\,$\micron$) range are significant for the M8 and M5 but relatively minor for the M2 star  (and negligible for earlier type stars not shown here). In the mid-infrared, stellar spectral lines are generally weaker than in the visible or near-infrared due to the lower temperature sensitivity of the Planck function in the Rayleigh–Jeans regime. In contrast, planetary emission in the Wien regime exhibits stronger molecular features, making the simulated mid-IR flux ratio $F_\mathrm{p, sim} / F_\mathrm{s, sim}$ inherently low for cool stars, independent of missing opacities. Nevertheless, opacity sources differ from stellar models, along with different treatment of pressure broadening.
The SPHINX spectral synthesis framework \citep{Iyer2023} incorporates up-to-date opacity sources tailored for M-dwarf atmospheres, with molecular and atomic data drawn primarily from ExoMol, EXOPLINES, and related databases \citep{GharibNezhad2021, Freedman2014, Polyansky2018, Allard2016, Yurchenko2017}. It includes detailed treatments of molecular opacities such as H$_2$O, CH$_4$, CO, TiO, VO, and metal hydrides, as well as H$_2$–H$_2$/He collision-induced absorption and pressure-broadened atomic lines. In contrast, the PHOENIX/BT-Settl models \citep{BT-Settl} rely on more classical opacity sources, including TiO from \citet{Plez1998}, H$_2$O from the BT2 line list \citep{Barber2006}, CO from \citet{Goorvitch1994}, and atomic data from Kurucz \citep{Kurucz_2005}, VALD, and NIST. While both models aim to capture key opacity features of cool stars, SPHINX emphasizes modern, high-temperature laboratory and theoretical data with a correlated-k approach to handle opacity sampling.

The bottom row of Figure\,\ref{Fig:specData} shows the difference in the simulated secondary eclipse depth ($F_\mathrm{p, sim} / F_\mathrm{s, sim}$) using either SPHINX or PHOENIX/BT-Settl stellar spectral models for $F_\mathrm{s, sim}$, with $F_\mathrm{p, sim}$ corresponding to the TRAPPIST-1\,b flux assuming a 490\,K blackbody \citep{ducrot_2024}. To isolate $F_\mathrm{s, sim}$ sensitivity, we fix $F_\mathrm{p, sim}$ regardless of $F_\mathrm{s, sim}$ value. Such differences reach 60\,ppm and 20\,ppm  at 15\,$\micron$ for the M8 and M5, respectively, while earlier type stars show great consistency between the two models. These values are beyond the precision requirement previously introduced and should thus impact the derived planetary properties beyond acceptable levels. 

\begin{figure}[t]
\centering 
\includegraphics[width=0.9\linewidth]{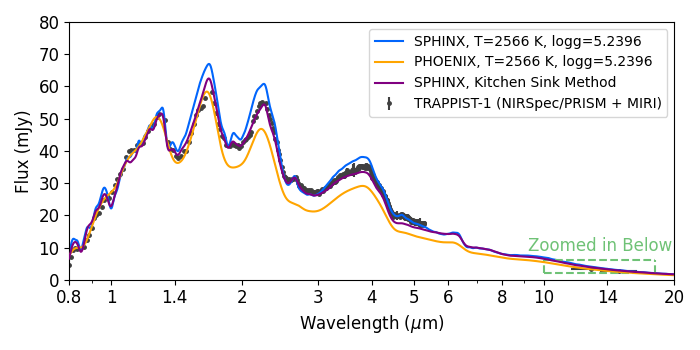}
\includegraphics[width=0.9\linewidth]{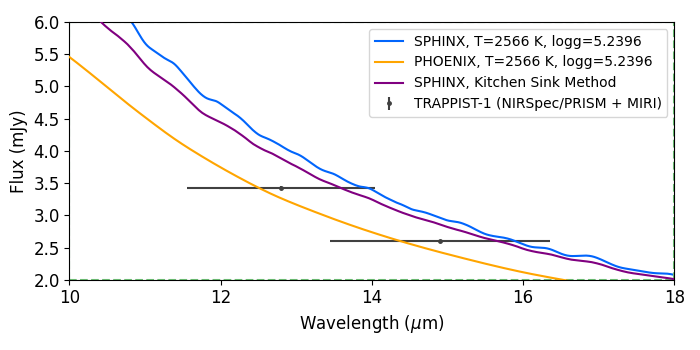}
\caption{{\bf Top:} Calibrated NIRSpec/PRISM spectrum of TRAPPIST-1 (black points) obtained from a single visit using publicly-available data (GTO-1331). We also include the MIRI 12.8 and 15\,$\micron$ photometric values \citep{ducrot_2024}.  SPHINX and PHOENIX models (colored lines) are normalized to the J-band (1.2 -- 1.3\,$\micron$) region of the NIRSpec/PRISM data.  The 2566~K SPHINX and PHOENIX models over- and under-predict, respectively, the NIRSpec and MIRI data longer than 1.3\,$\micron$.  The Kitchen Sink method provides the best comparison; however, still over-predicts the MIRI data.  
{\bf Bottom}: Zoomed in view from 10 -- 18\,$\micron$. The calibrated MIRI photometric points of 3.426 $\pm$ 0.12 mJy at 12.8\,$\micron$ and 2.543 $\pm$ 0.097 mJy at 15.0\,$\micron$ \citep{ducrot_2024}, with error bars too small to be seen, are well-below the SPHINX models and well-above the PHOENIX model.  This demonstrates that extrapolating calibrated TRAPPIST-1 stellar models from NIR to MIR wavelengths can yield unreliable predictions.
}
\label{Fig:NIRSpec}
\vspace*{-1.\baselineskip}
\end{figure}

\section{Discussion}\label{sec:discussion}

``Know Thy Star, Know Thy Planet'' has long been the mantra of the exoplanet detection and characterization communities for good reason. For the exoplanet characterization community, this statement highlights the TLS effect \citep{Rackham2018, Rackham2023} impacting JWST transit studies \citep{Lim_2023, May_2023, Moran2023, Cadieux_2024, Rathcke2025}. Emission spectroscopy in secondary eclipse offers a powerful way to characterize warm-to-hot transiting exoplanets for which the TLS effect might hamper characterization in transmission. Indeed, contrary to transmission spectroscopy, even the absence of an atmosphere does not prohibit characterization via emission, as surface properties, such as albedo and mineralogy \citep{First_2024,Paragas_2025}, can be constrained from eclipse depth measurements. But, to date, the interpretation of the eclipse depth measurements to derive the planet surface and/or atmospheric properties rely on choosing a synthetic stellar mid-IR spectrum model among a variety of options; empirical constraints on MIR stellar spectra are limited in the pre-JWST era \citep[e.g.,][]{Ardila2010}. In this context, we argue here that ``Know Thy Star, Know Thy Planet'' holds also for secondary eclipse observation of M-dwarf planets, given the biases introduced by differences in those stellar model spectra compared to real stellar flux measurements.

\subsection{Truth May Be Further Than We Think}\label{subsec:accuracy}
\cite{ducrot_2024} have estimated that a 29~ppm precision in the 15\,$\micron$ eclipse depth would be required to measure surface albedo with a precision of 0.1 at 3$\sigma$ confidence, while the current 3$\sigma$ measurement uncertainty is 270~ppm. 
We show here that the choice of the stellar model needed to derive the 15\,$\micron$ eclipse depth already introduces a 40--80~ppm uncertainty, therefore pushing the measurement beyond the precision requirement. Yet, even such model uncertainty is not an estimation of the accuracy of those synthetic spectra with respect to a true, measured mid-IR spectrum. 

Indeed, using the 2569~$\pm 28$\,K effective temperature derived from NIRSpec/PRISM measurements \citep{Davoudi_2024}, the J-band calibrated models  significantly over-predict the measured flux in the MIRI range at 12.8\,$\micron$ and 15\,$\micron$ \citep{Greene_2023,Zieba_2023, ducrot_2024} by about  13$\%$ for PHOENIX and about 7$\%$ for SPHINX. However, \citet{Greene_2023} employed an out-dated version of PHOENIX/BT-Settl produced by the pysynphot package that preceded the last update in 2021 for M/H=0.0 (T.\ Greene, private communication). With the last available version of the model, we now estimate a 2$\%$ flux discrepancy at 15\,$\micron$. \cite{ducrot_2024} followed a similar approach to \cite{Ih2023}, scaling the SPHINX stellar spectrum model to the measured flux value (applying a 7$\%$ correction) and using this corrected version to model the planetary surfaces and atmospheres discussed in their study. More recently, \cite{Valdes_2025} adopted a comparable method for the case of TOI-1468 b. This difference exemplifies that the use of synthetic spectra rather than the real measured spectrum is subject to errors and version differences, impacting the scientific results.

Instead of calibrating to the J-band magnitude \citep{Greene_2023}, we explore calibrating models to the J-band region (1.2 -- 1.3\,$\micron$) of the NIRSpec/PRISM data shown in Figure \ref{Fig:NIRSpec}.  Note that the NIRSpec data are 4.0 mJy lower than predicted for a J=11.35 object (48.8 vs 52.8 mJy). The 2566~K SPHINX model slightly over-predicts the NIRSpec data longer than 1.3\,$\micron$ and drastically over-predicts the two MIRI broadband points.  Conversely, the 2566~K PHOENIX model drastically under-predicts the NIRSpec data longer than 1.3\,$\micron$ and slightly under-predicts the two MIRI broadband points contrary to the over-prediction in \cite{Greene_2023}. In hopes of achieving a better match to the calibrated JWST spectrum, we compute and display a third model in Figure \ref{Fig:NIRSpec} that combines multiple 1D stellar components. We call this the ``Kitchen Sink'' method and describe it below.


Disk heterogeneity is typical for active stars like TRAPPIST-1 and, inevitably, a single one-dimensional representation of the star could fail to capture heterogeneous details, particularly if the variations are large in temperature or surface area. To combat this, other works have utilized multi-component fits to the TRAPPIST-1 spectrum, which included additional components to represent spots and faculae \citep[e.g,][]{Zhang2018, Wakeford2019, Garcia2022, Davoudi_2024}. From the Sun, we understand that spot temperatures and sizes vary across the lifetime of a spot and thus take this one step further. Using the JWST NIRSpec/PRISM data, we employ a linear combination of hundreds of spectral components from the SPHINX stellar grid in a range of T$_\mathrm{eff}$=\{2000,4000\}\,K, $\log$\,g=\{4.75,5.25\}, [M/H]=\{-0.25,0.5\}, and C/O=\{0.3,0.9\} to achieve a more robust fit than previous works (1008 models). Using a $\chi^2$ minimizer, we fit for the weight of each spectrum and the radius of the star, keeping the distance to the star fixed and initialized to 100\% weight in the 2600\,K, $\log$\,g=5.25, [M/H]=0, C/O=0.3 spectrum.

Our best-fit model is composed of predominantly 2400--2700\,K models with metallicities ranging from 0.25--0.5, gravities ranging from 4.75--5.25, and C/O ratios of either 0.3 or 0.9. The top eight model components determine 95\% of the best-fit solution.  Residuals to the spectrum are typically less than a few mJy, with the exception of the region near 1.7\,$\mu$m. This corresponds to a deviation of the residuals of approximately \textless10\% from 1--5 $\mu$m.  Even with this optimized combination of SPHINX models, the extrapolated fluxes at 12.8 and 15\,$\mu$m continue to exceed the observed values (Figure \ref{Fig:NIRSpec}, bottom panel).
Such discrepancy strongly suggests that unknown features of the star (i.e., spots, silicate clouds, molecular absorption) reduce the mid-IR stellar flux, at least in the 12.8\,$\micron$ and 15\,$\micron$ channels. 

The biases in the current modeling of stellar mid-IR spectra may therefore be more significant than what we presently perceive as the differences between models. This raises a critical question: if the deviation from ``truth'' is already substantial for TRAPPIST-1, how far off might our models may be for other stars? The extent of these biases remains unclear, emphasizing the urgent need for a thorough reassessment of observational strategies.

\subsection{From Albedo to Spectra}\label{subsec:albedo}
Recently, \cite{Paragas_2025} have shown using laboratory measurements that albedo is a poor proxy for surface types. Indeed, slight changes in the surface texture can significantly alter its albedo. Therefore, these authors recommend the use of spectrally resolved data to constrain surface types for exoplanet observations. Coincidentally, JWST Guest Observer program GO 6219 (PI Achrene Dyrek) aims at testing the use of MIRI LRS in fixed-slit mode to maximize the signal-to-noise ratio of the secondary eclipse observations. For cool stars like TRAPPIST-1, it would be particularly efficient and could help to significantly reduce the number of eclipses needed. The spectral information that would be brought by a MIRI LRS mid-IR spectrum in a secondary eclipse would potentially help to lift the high degeneracy associate with the inference of possible surface types \citep{ducrot_2024} and could be compared to the current photometric data points at 12.8 and 15\,$\micron$. 

\subsection{A Mitigation Strategy to Implement Now}\label{subsec:mitigation}

At the dawn of large DDT programs dedicated to hot rocky exoplanets, such as the 500-hour Rocky Worlds DDT initiative and the ``The Hot Rocks Survey'' (Cycle 2 proposal 3730, PI Diamond-Lowe), it is crucial to mitigate biases in the interpretation of these data. Rocky Worlds specifically calls the community to identify ancillary data which would facilitate the data interpretation. 
Here we argue for the systematic collection of MIRI MRS mid-IR stellar spectra for every secondary eclipse measurement campaign to avoid the use of biased stellar models. Note that while it is possible to obtain the true $F_p$ in a given filter using MIRI Imaging, the challenge arises when plotting the emission spectrum and comparing broadband measurements to spectral curves extending over longer wavelengths. Simply scaling the synthetic stellar spectrum to a single stellar flux measurement in a broadband filter may not be sufficient. Since these stars are not well characterized, there may be unresolved spectral features in the mid-IR region that need to be properly accounted for when generating emission models (atmosphere, surface and even theoretical blackbodies curves) for comparison with the data. If a MIRI MRS spectrum of the star is available, it becomes unnecessary to interpret the planetary flux $F_\mathrm{p, obs}$ using isolated photometric bands (e.g., 12.8 or 15\,$\micron$) and to scale a stellar model based on one or two photometric points, which may yield inconsistent scaling factors. There is no physical justification for assuming that the true stellar spectrum is simply proportional to any given model. Instead, the observed planetary spectrum could be directly compared to synthetic planetary spectra for surface inference, using the measured stellar spectrum as a reference—analogous to the approach illustrated in Figure \ref{Fig:eclipse}, but without relying on synthetic stellar spectra.

For instance, \citep{Miles_2023} found from JWST MIRI observations of the brown dwarf VHS 1256 b that water vapor is present between 10 to 16\,$\micron$ and silicate clouds between 8 and 12\,$\micron$. TRAPPIST-1 is very close to the brown dwarf transition and its lower than expected fluxes at 12.8 and 15\,$\micron$ may also be caused by water vapor and silicate clouds. While both SPHINX and PHOENIX/BT-Settl \citep{BT-Settl} contain water vapor absorption, enhanced H$_2$O vapor beyond the models is a possible factor, though it might not fully account for the average of 10\% discrepancy that we observe. However, a lower $\log g$ than 5.2 \citep{Agol2021} and/or metallicity $\mathrm{[Fe/H]} > 0$, as shown in Figure 2, could also contribute to reducing the stellar flux in the mid-IR. Moreover, such late M dwarfs could also exhibit surface inhomogeneities such as cool spots, clouds, or dynamic structures that are not fully captured by 1D models. Lastly,  observational systematics (e.g., calibration errors in the mid-IR photometry)---especially in the thermal IR, where sky backgrounds and detector behavior are trickier---could systematically lead to lower fluxes than what stellar models predict. Observing a true mid-IR spectrum at moderate resolution would help to disentangle these possible effects.

For TRAPPIST-1, using MIRI MRS mode (4.9 to 27.9\,$\micron$) at $R{=}5800$, we estimate that a minimum SNR of 50 per resolution element (150 at 10\,$\micron$ and 80 at 15\,$\micron$), enabling sufficient resolution of most of the expected mid-IR spectral lines, would be acquired in just 2.57 hours (overheads included) of continuous observation of the star, making such an essential ancillary observation relatively inexpensive. However, for some targets it may be recommended to acquire such spectrum at several epochs to test the mid-IR temporal stability hypothesis. If the variability is minimal, then a single mid-IR spectrum per target would be enough.

We also argue that the medium  resolution of MRS should be favored over the lower resolution of LRS to avoid blending absorption lines that remain to be detected. However, MIRI LRS could serve as a valuable upgrade to  broadband photometry, especially if the approved program (GO 6219, PI Dyrek) successfully demonstrates the efficiency gains of MIRI LRS fixed-slit.
A complete MIRI MRS stellar spectrum from 4.9 to 27.9\,$\micron$ would diagnose whether the mid-IR flux is globally lower than suggested by the (NIR-derived) effective temperature, therefore implying that such star cannot be fit with a single stellar model from the UV to the mid-IR, or that the reduction in flux is spectrally local, due to some specific absorbers that need to be characterized.
Note that acquiring such a calibrated mid-IR stellar spectrum would provide a direct continuity from the the visible and NIR spectra that are now required to correct for TLS effects. Extending the spectral energy distribution fitting across a broader and more precise wavelength range would also enable refinements of the stellar parameters, such as effective temperature, $\log g$, and metallicity, and would help mitigate uncertainties associated with current SED flux measurements \citep{Davoudi_2024}.

\section{Conclusions}\label{sec:conclusions}
In conclusion, we advocate for using JWST's calibrated stellar flux when characterizing rocky exoplanets in secondary eclipse. It has been shown for mid to late M dwarf that the inferred planet flux can be biased by more than 20\%, relative to the measured stellar flux, depending on which stellar model  is used to fit the calibrated NIRSpec data and then extrapolated out to 12.8 or 15\,$\micron$.  These inconsistencies in the stellar flux lead to large discrepancies in the retrieved planetary albedo or bare surface type and could result in the false detection of an atmosphere.

The significant differences between the stellar models and the measured flux at 12.8 and 15\,$\micron$ points to missing physics in our models that cannot be identified using broadband photometry. Furthermore, we show that discrepancies between versions of the same stellar model could significantly affect the difference between the model and measured fluxes in the MIRI bandpasses. Thus, we advocate for acquiring mid-IR spectra with MIRI MRS as ancillary data to MIRI photometry observations.  Acquiring MRS data is relatively inexpensive. For TRAPPIST-1 we estimate that 2.6\,hr (overheads included) should be sufficient to obtain a $R{=}5800$, 5--20\,$\micron$ spectrum, leading to a SNR of 80 at 15\,$\micron$.  We also recommend that such ancillary data be considered in the 500-hour Rocky Worlds DDT initiative.

A mid-IR spectrum of the star can be used to assess whether photometric data points align with the mid-IR stellar spectrum. If they do not,  the detailed spectral information provided by MRS would provide a unique means to highlight and review potential calibration issues or indications of stellar mid-IR variability (e.g., flares, spots, silicate clouds, or molecular features). Furthermore, multiple mid-IR spectra taken at various epochs could provide critical insights on the mid-IR variability of a given star, which can only be sporadically understood via broadband photometry.

By consistently integrating mid-IR stellar observations into exoplanet characterization efforts, we not only enhance our ability to accurately interpret exoplanetary atmospheres, but also establish a robust framework for evaluating stellar models across a wide range of stellar types. This approach will lead to more precise and unbiased interpretations of both stellar and exoplanetary spectra, ultimately advancing our understanding of planetary environments beyond our solar system.

\section{acknowledgments}
TJF, KBS, and LCM acknowledge support from the CHAMPs (Consortium on Habitability and Atmospheres of M-dwarf Planets) team, supported by the National Aeronautics and Space Administration (NASA) under grant no.\ 80NSSC23K1399 issued through the Interdisciplinary Consortia for Astrobiology Research (ICAR) program.
B.V.R. and J.d.W. acknowledge support from the European Research Council (ERC) Synergy Grant under the European Union’s Horizon 2020 research and innovation program (grant No. 101118581 — project REVEAL).
This material is based upon work supported by the National Aeronautics and Space Administration under Agreement No.\ 80NSSC21K0593 for the program ``Alien Earths''.
The results reported herein benefited from collaborations and/or information exchange within NASA’s Nexus for Exoplanet System Science (NExSS) research coordination network sponsored by NASA’s Science Mission Directorate.
The authors thank the two anonymous referees for their careful review of our manuscript. The authors also thank Sasha Shapiro for enlightening feedback on MIR stellar spectral features, Tom Greene for their help on the PHOENIX/BT-Setll version and Martin Turbet and Geronimo Villanueva for their useful feedback on this paper.








\bibliography{main}{}

\begin{thebibliography}{}
\expandafter\ifx\csname natexlab\endcsname\relax\def\natexlab#1{#1}\fi
\providecommand{\url}[1]{\href{#1}{#1}}
\providecommand{\dodoi}[1]{doi:~\href{http://doi.org/#1}{\nolinkurl{#1}}}
\providecommand{\doeprint}[1]{\href{http://ascl.net/#1}{\nolinkurl{http://ascl.net/#1}}}
\providecommand{\doarXiv}[1]{\href{https://arxiv.org/abs/#1}{\nolinkurl{https://arxiv.org/abs/#1}}}

\bibitem[{{Agol} {et~al.}(2021){Agol}, {Dorn}, {Grimm}, {Turbet}, {Ducrot}, {Delrez}, {Gillon}, {Demory}, {Burdanov}, {Barkaoui}, {Benkhaldoun}, {Bolmont}, {Burgasser}, {Carey}, {de Wit}, {Fabrycky}, {Foreman-Mackey}, {Haldemann}, {Hernandez}, {Ingalls}, {Jehin}, {Langford}, {Leconte}, {Lederer}, {Luger}, {Malhotra}, {Meadows}, {Morris}, {Pozuelos}, {Queloz}, {Raymond}, {Selsis}, {Sestovic}, {Triaud}, \& {Van Grootel}}]{Agol2021}
{Agol}, E., {Dorn}, C., {Grimm}, S.~L., {et~al.} 2021, PSJ, 2, 1, \dodoi{10.3847/PSJ/abd022}

\bibitem[{{Allard}(2014)}]{BT-Settl}
{Allard}, F. 2014, in Exploring the Formation and Evolution of Planetary Systems, ed. M.~{Booth}, B.~C. {Matthews}, \& J.~R. {Graham}, Vol. 299, 271--272, \dodoi{10.1017/S1743921313008545}

\bibitem[{Allard {et~al.}(2016)Allard, Allard, Homeier, Freytag, Rajpurohit, \& Schultheis}]{Allard2016}
Allard, N.~F., Allard, F., Homeier, D., {et~al.} 2016, Astronomy \& Astrophysics, 588, A142, \dodoi{10.1051/0004-6361/201527595}

\bibitem[{{Ardila} {et~al.}(2010){Ardila}, {Van Dyk}, {Makowiecki}, {Stauffer}, {Song}, {Rho}, {Fajardo-Acosta}, {Hoard}, \& {Wachter}}]{Ardila2010}
{Ardila}, D.~R., {Van Dyk}, S.~D., {Makowiecki}, W., {et~al.} 2010, \apjs, 191, 301, \dodoi{10.1088/0067-0049/191/2/301}

\bibitem[{August {et~al.}(2024)August, Buchhave, Diamond-Lowe, Mendonça, Gressier, Rathcke, Allen, Fortune, Jones, Meier-Valdés, Demory, Espinoza, Fisher, Gibson, Heng, Hoeijmakers, Hooton, Kitzmann, \& Prinoth}]{August_2024}
August, P.~C., Buchhave, L.~A., Diamond-Lowe, H., {et~al.} 2024, Hot Rocks Survey I : A shallow eclipse for LHS 1478 b.
\newblock \doarXiv{2410.11048}

\bibitem[{Barber {et~al.}(2006)Barber, Tennyson, Harris, \& Tolchenov}]{Barber2006}
Barber, R.~J., Tennyson, J., Harris, G.~J., \& Tolchenov, R.~N. 2006, Monthly Notices of the Royal Astronomical Society, 368, 1087, \dodoi{10.1111/j.1365-2966.2006.10184.x}

\bibitem[{{Berardo} {et~al.}(2024){Berardo}, {de Wit}, \& {Rackham}}]{Berardo2024}
{Berardo}, D., {de Wit}, J., \& {Rackham}, B.~V. 2024, \apjl, 961, L18, \dodoi{10.3847/2041-8213/ad1b5b}

\bibitem[{Cadieux {et~al.}(2024)Cadieux, Doyon, MacDonald, Turbet, Artigau, Lim, Radica, Fauchez, Salhi, Dang, Albert, Coulombe, Cowan, LafreniÃšre, LâHeureux, Piaulet-Ghorayeb, Benneke, Cloutier, Charnay, Cook, Fournier-Tondreau, Plotnykov, \& Valencia}]{Cadieux_2024}
Cadieux, C., Doyon, R., MacDonald, R.~J., {et~al.} 2024, The Astrophysical Journal Letters, 970, L2, \dodoi{10.3847/2041-8213/ad5afa}

\bibitem[{Coulombe {et~al.}(2023)Coulombe, Benneke, Challener, {et~al.}}]{Coulombe_2023}
Coulombe, L., Benneke, B., Challener, R., {et~al.} 2023, Nature, 620, 292, \dodoi{10.1038/s41586-023-06230-1}

\bibitem[{Davoudi {et~al.}(2024)Davoudi, Rackham, Gillon, de~Wit, Burgasser, Delrez, Iyer, \& Ducrot}]{Davoudi_2024}
Davoudi, F., Rackham, B.~V., Gillon, M., {et~al.} 2024, The Astrophysical Journal Letters, 970, L4, \dodoi{10.3847/2041-8213/ad5c6c}

\bibitem[{{Ducrot} {et~al.}(2024){Ducrot}, {Lagage}, {Min}, {Gillon}, {Bell}, {Tremblin}, {Greene}, {Dyrek}, {Bouwman}, {Waters}, {G{\"u}del}, {Henning}, {Vandenbussche}, {Absil}, {Barrado}, {Boccaletti}, {Coulais}, {Decin}, {Edwards}, {Gastaud}, {Glasse}, {Kendrew}, {Olofsson}, {Patapis}, {Pye}, {Rouan}, {Whiteford}, {Argyriou}, {Cossou}, {Glauser}, {Krause}, {Lahuis}, {Royer}, {Scheithauer}, {Colina}, {van Dishoeck}, {Ostlin}, {Ray}, \& {Wright}}]{ducrot_2024}
{Ducrot}, E., {Lagage}, P.-O., {Min}, M., {et~al.} 2024, Nature Astronomy, \dodoi{10.1038/s41550-024-02428-z}

\bibitem[{{Espinoza} {et~al.}(2019){Espinoza}, {Rackham}, {Jord{\'a}n}, {Apai}, {L{\'o}pez-Morales}, {Osip}, {Grimm}, {Hoeijmakers}, {Wilson}, {Bixel}, {McGruder}, {Rodler}, {Weaver}, {Lewis}, {Fortney}, \& {Fraine}}]{Espinoza2019}
{Espinoza}, N., {Rackham}, B.~V., {Jord{\'a}n}, A., {et~al.} 2019, \mnras, 482, 2065, \dodoi{10.1093/mnras/sty2691}

\bibitem[{First {et~al.}(2024)First, Mishra, Gazel, {et~al.}}]{First_2024}
First, E., Mishra, I., Gazel, E., {et~al.} 2024, Nature Astronomy, \dodoi{10.1038/s41550-024-02412-7}

\bibitem[{Freedman {et~al.}(2014)Freedman, Lustig-Yaeger, Fortney, Lupu, Marley, \& Lodders}]{Freedman2014}
Freedman, R.~S., Lustig-Yaeger, J., Fortney, J.~J., {et~al.} 2014, The Astrophysical Journal Supplement Series, 214, 25, \dodoi{10.1088/0067-0049/214/2/25}

\bibitem[{{Garcia} {et~al.}(2022){Garcia}, {Moran}, {Rackham}, {Wakeford}, {Gillon}, {de Wit}, \& {Lewis}}]{Garcia2022}
{Garcia}, L.~J., {Moran}, S.~E., {Rackham}, B.~V., {et~al.} 2022, \aap, 665, A19, \dodoi{10.1051/0004-6361/202142603}

\bibitem[{Gharib-Nezhad \& Line(2021)}]{GharibNezhad2021}
Gharib-Nezhad, E., \& Line, M.~R. 2021, The Astrophysical Journal, 910, 90, \dodoi{10.3847/1538-4357/abe38e}

\bibitem[{Gillon {et~al.}(2016)Gillon, Jehin, Lederer, Delrez, de~Wit, Burdanov, Van~Grootel, Burgasser, Triaud, Opitom, Demory, Sahu, Bardalez~Gagliuffi, Magain, \& Queloz}]{Gillon2016}
Gillon, M., Jehin, E., Lederer, S.~M., {et~al.} 2016, Nature, 533, 221 .
\newblock \url{https://doi.org/10.1038/nature17448}

\bibitem[{Gillon {et~al.}(2017)Gillon, Triaud, Demory, Jehin, Agol, Deck, Lederer, de~Wit, Burdanov, Ingalls, Bolmont, Leconte, Raymond, Selsis, Turbet, Barkaoui, Burgasser, Burleigh, Carey, Chaushev, Copperwheat, Delrez, Fernandes, Holdsworth, Kotze, Van~Grootel, Almleaky, Benkhaldoun, Magain, \& Queloz}]{Gillon2017}
Gillon, M., Triaud, A. H. M.~J., Demory, B.-O., {et~al.} 2017, Nature, 542, 456–460.
\newblock \url{https://doi.org/10.1038/nature21360}

\bibitem[{Goorvitch(1994)}]{Goorvitch1994}
Goorvitch, D. 1994, The Astrophysical Journal Supplement Series, 95, 535, \dodoi{10.1086/192110}

\bibitem[{Greene {et~al.}(2023)Greene, Bell, Ducrot, Dyrek, Lagage, \& Fortney}]{Greene_2023}
Greene, T.~P., Bell, T.~J., Ducrot, E., {et~al.} 2023, Nature, 618, 39, \dodoi{10.1038/s41586-023-05951-7}

\bibitem[{{Ih} {et~al.}(2023){Ih}, {Kempton}, {Whittaker}, \& {Lessard}}]{Ih2023}
{Ih}, J., {Kempton}, E. M.~R., {Whittaker}, E.~A., \& {Lessard}, M. 2023, \apjl, 952, L4, \dodoi{10.3847/2041-8213/ace03b}

\bibitem[{{Iyer} \& {Line}(2020)}]{Iyer2020}
{Iyer}, A.~R., \& {Line}, M.~R. 2020, \apj, 889, 78, \dodoi{10.3847/1538-4357/ab612e}

\bibitem[{{Iyer} {et~al.}(2023){Iyer}, {Line}, {Muirhead}, {Fortney}, \& {Gharib-Nezhad}}]{Iyer2023}
{Iyer}, A.~R., {Line}, M.~R., {Muirhead}, P.~S., {Fortney}, J.~J., \& {Gharib-Nezhad}, E. 2023, \apj, 944, 41, \dodoi{10.3847/1538-4357/acabc2}

\bibitem[{Iyer {et~al.}(2022)Iyer, Line, Muirhead, Fortney, \& Gharib-Nezhad}]{SPHINX}
Iyer, R.~A., Line, R.~M., Muirhead, S.~P., Fortney, J.~J., \& Gharib-Nezhad, E. 2022, \dodoi{10.5281/zenodo.7416042}

\bibitem[{{Kirk} {et~al.}(2021){Kirk}, {Rackham}, {MacDonald}, {L{\'o}pez-Morales}, {Espinoza}, {Lendl}, {Wilson}, {Osip}, {Wheatley}, {Skillen}, {Apai}, {Bixel}, {Gibson}, {Jord{\'a}n}, {Lewis}, {Louden}, {McGruder}, {Nikolov}, {Rodler}, \& {Weaver}}]{Kirk2021}
{Kirk}, J., {Rackham}, B.~V., {MacDonald}, R.~J., {et~al.} 2021, \aj, 162, 34, \dodoi{10.3847/1538-3881/abfcd2}

\bibitem[{Koll(2022)}]{Koll_2022}
Koll, D. D.~B. 2022, The Astrophysical Journal, 924, 134, \dodoi{10.3847/1538-4357/ac3b48}

\bibitem[{{Krissansen-Totton} \& {Fortney}(2022)}]{Krissansen-Totton_2022}
{Krissansen-Totton}, J., \& {Fortney}, J.~J. 2022, \apj, 933, 115, \dodoi{10.3847/1538-4357/ac69cb}

\bibitem[{{Kurucz}(2005)}]{Kurucz_2005}
{Kurucz}, R.~L. 2005, Memorie della Societa Astronomica Italiana Supplementi, 8, 14

\bibitem[{{Lim} {et~al.}(2023){Lim}, {Benneke}, {Doyon}, {MacDonald}, {Piaulet}, {Artigau}, {Coulombe}, {Radica}, {L'Heureux}, {Albert}, {Rackham}, {de Wit}, {Salhi}, {Roy}, {Flagg}, {Fournier-Tondreau}, {Taylor}, {Cook}, {Lafreni{\`e}re}, {Cowan}, {Kaltenegger}, {Rowe}, {Espinoza}, {Dang}, \& {Darveau-Bernier}}]{Lim_2023}
{Lim}, O., {Benneke}, B., {Doyon}, R., {et~al.} 2023, \apjl, 955, L22, \dodoi{10.3847/2041-8213/acf7c4}

\bibitem[{{Lincowski} {et~al.}(2018){Lincowski}, {Meadows}, {Crisp}, {Robinson}, {Luger}, {Lustig-Yaeger}, \& {Arney}}]{Lincowski2018}
{Lincowski}, A.~P., {Meadows}, V.~S., {Crisp}, D., {et~al.} 2018, The Astrophysical Journal, 867, 76, \dodoi{10.3847/1538-4357/aae36a}

\bibitem[{{Luger} {et~al.}(2017){Luger}, {Sestovic}, {Kruse}, {Grimm}, {Demory}, {Agol}, {Bolmont}, {Fabrycky}, {Fernandes}, {Van Grootel}, {Burgasser}, {Gillon}, {Ingalls}, {Jehin}, {Raymond}, {Selsis}, {Triaud}, {Barclay}, {Barentsen}, {Howell}, {Delrez}, {de Wit}, {Foreman-Mackey}, {Holdsworth}, {Leconte}, {Lederer}, {Turbet}, {Almleaky}, {Benkhaldoun}, {Magain}, {Morris}, {Heng}, \& {Queloz}}]{Luger2017}
{Luger}, R., {Sestovic}, M., {Kruse}, E., {et~al.} 2017, Nature Astronomy, 1, 0129, \dodoi{10.1038/s41550-017-0129}

\bibitem[{{Marley} {et~al.}(1999){Marley}, {Gelino}, {Stephens}, {Lunine}, \& {Freedman}}]{Marley1999}
{Marley}, M.~S., {Gelino}, C., {Stephens}, D., {Lunine}, J.~I., \& {Freedman}, R. 1999, \apj, 513, 879, \dodoi{10.1086/306881}

\bibitem[{May {et~al.}(2023)May, MacDonald, Bennett, Moran, Wakeford, Peacock, Lustig-Yaeger, Highland, Stevenson, Sing, Mayorga, Batalha, Kirk, LÃ³pez-Morales, Valenti, Alam, Alderson, Fu, Gonzalez-Quiles, Lothringer, Rustamkulov, \& Sotzen}]{May_2023}
May, E.~M., MacDonald, R.~J., Bennett, K.~A., {et~al.} 2023, The Astrophysical Journal Letters, 959, L9, \dodoi{10.3847/2041-8213/ad054f}

\bibitem[{{McCullough} {et~al.}(2014){McCullough}, {Crouzet}, {Deming}, \& {Madhusudhan}}]{McCullough2014}
{McCullough}, P.~R., {Crouzet}, N., {Deming}, D., \& {Madhusudhan}, N. 2014, \apj, 791, 55, \dodoi{10.1088/0004-637X/791/1/55}

\bibitem[{{Meier Valdes, E. A.} {et~al.}(2025){Meier Valdes, E. A.}, {Demory, B.-O.}, {Diamond-Lowe, H.}, {Mendonza, J. M.}, {August, P. C.}, {Fortune, M.}, {Allen, N. H.}, {Kitzmann, D.}, {Gressier, A.}, {Hooton, M.}, {Jones, K. D.}, {Buchhave, L. A.}, {Espinoza, N.}, {Fisher, C. E.}, {Gibson, N. P.}, {Heng, K.}, {Hoeijmakers, J.}, {Prinoth, B.}, {Rathcke, A. D.}, \& {Eastman, J. D.}}]{Valdes_2025}
{Meier Valdes, E. A.}, {Demory, B.-O.}, {Diamond-Lowe, H.}, {et~al.} 2025, A\&A, 698, A68, \dodoi{10.1051/0004-6361/202453449}

\bibitem[{Miles {et~al.}(2023)Miles, Biller, Patapis, Worthen, Rickman, Hoch, Skemer, Perrin, Whiteford, Chen, Sargent, Mukherjee, Morley, Moran, Bonnefoy, Petrus, Carter, Choquet, Hinkley, Ward-Duong, Leisenring, Millar-Blanchaer, Pueyo, Ray, Sallum, Stapelfeldt, Stone, Wang, Absil, Balmer, Boccaletti, Bonavita, Booth, Bowler, Chauvin, Christiaens, Currie, Danielski, Fortney, Girard, Grady, Greenbaum, Henning, Hines, Janson, Kalas, Kammerer, Kennedy, Kenworthy, Kervella, Lagage, Lew, Liu, Macintosh, Marino, Marley, Marois, Matthews, Matthews, Mawet, McElwain, Metchev, Meyer, Molliere, Pantin, Quirrenbach, Rebollido, Ren, Schneider, Vasist, Wyatt, Zhou, Briesemeister, Bryan, Calissendorff, Cantalloube, Cugno, Furio, Dupuy, Factor, Faherty, Fitzgerald, Franson, Gonzales, Hood, Howe, Kraus, Kuzuhara, Lagrange, Lawson, Lazzoni, Liu, Llop-Sayson, Lloyd, Martinez, Mazoyer, Quanz, Redai, Samland, Schlieder, Tamura, Tan, Uyama, Vigan, Vos, Wagner, Wolff, Ygouf, Zhang, Zhang, \& Zhang}]{Miles_2023}
Miles, B.~E., Biller, B.~A., Patapis, P., {et~al.} 2023, The Astrophysical Journal Letters, 946, L6, \dodoi{10.3847/2041-8213/acb04a}

\bibitem[{{Moran} {et~al.}(2023){Moran}, {Stevenson}, {Sing}, {MacDonald}, {Kirk}, {Lustig-Yaeger}, {Peacock}, {Mayorga}, {Bennett}, {L{\'o}pez-Morales}, {May}, {Rustamkulov}, {Valenti}, {Adams Redai}, {Alam}, {Batalha}, {Fu}, {Gonzalez-Quiles}, {Highland}, {Kruse}, {Lothringer}, {Ortiz Ceballos}, {Sotzen}, \& {Wakeford}}]{Moran2023}
{Moran}, S.~E., {Stevenson}, K.~B., {Sing}, D.~K., {et~al.} 2023, \apjl, 948, L11, \dodoi{10.3847/2041-8213/accb9c}

\bibitem[{{Narrett} {et~al.}(2024){Narrett}, {Rackham}, \& {de Wit}}]{Narrett2024}
{Narrett}, I.~S., {Rackham}, B.~V., \& {de Wit}, J. 2024, \aj, 167, 107, \dodoi{10.3847/1538-3881/ad1f6c}

\bibitem[{Paragas {et~al.}(2025)Paragas, Knutson, Hu, Ehlmann, Alemanno, Helbert, Maturilli, Zhang, Iyer, \& Rossman}]{Paragas_2025}
Paragas, K., Knutson, H.~A., Hu, R., {et~al.} 2025, A New Spectral Library for Modeling the Surfaces of Hot, Rocky Exoplanets.
\newblock \doarXiv{2502.04433}

\bibitem[{{Park Coy} {et~al.}(2024){Park Coy}, {Ih}, {Kite}, {Koll}, {Tenthoff}, {Bean}, {Weiner Mansfield}, {Zhang}, {Xue}, {Kempton}, {Wolhfarth}, {Hu}, {Lyu}, \& {Wohler}}]{Coy_2024}
{Park Coy}, B., {Ih}, J., {Kite}, E.~S., {et~al.} 2024, arXiv e-prints, arXiv:2412.06573, \dodoi{10.48550/arXiv.2412.06573}

\bibitem[{Plez(1998)}]{Plez1998}
Plez, B. 1998, Astronomy and Astrophysics, 337, 495

\bibitem[{Polyansky {et~al.}(2018)Polyansky, Kyuberis, Zobov, Tennyson, Yurchenko, \& Lodi}]{Polyansky2018}
Polyansky, O.~L., Kyuberis, A.~A., Zobov, N.~F., {et~al.} 2018, Monthly Notices of the Royal Astronomical Society, 480, 2597, \dodoi{10.1093/mnras/sty1877}

\bibitem[{{Pont} {et~al.}(2008){Pont}, {Knutson}, {Gilliland}, {Moutou}, \& {Charbonneau}}]{Pont2008}
{Pont}, F., {Knutson}, H., {Gilliland}, R.~L., {Moutou}, C., \& {Charbonneau}, D. 2008, \mnras, 385, 109, \dodoi{10.1111/j.1365-2966.2008.12852.x}

\bibitem[{{Rackham} {et~al.}(2017){Rackham}, {Espinoza}, {Apai}, {L{\'o}pez-Morales}, {Jord{\'a}n}, {Osip}, {Lewis}, {Rodler}, {Fraine}, {Morley}, \& {Fortney}}]{Rackham2017}
{Rackham}, B., {Espinoza}, N., {Apai}, D., {et~al.} 2017, \apj, 834, 151, \dodoi{10.3847/1538-4357/aa4f6c}

\bibitem[{{Rackham} {et~al.}(2018){Rackham}, {Apai}, \& {Giampapa}}]{Rackham2018}
{Rackham}, B.~V., {Apai}, D., \& {Giampapa}, M.~S. 2018, \apj, 853, 122, \dodoi{10.3847/1538-4357/aaa08c}

\bibitem[{{Rackham} {et~al.}(2019){Rackham}, {Apai}, \& {Giampapa}}]{Rackham2019}
---. 2019, \aj, 157, 96, \dodoi{10.3847/1538-3881/aaf892}

\bibitem[{{Rackham} \& {de Wit}(2024)}]{Rackham2024}
{Rackham}, B.~V., \& {de Wit}, J. 2024, \aj, 168, 82, \dodoi{10.3847/1538-3881/ad5833}

\bibitem[{{Rackham} {et~al.}(2023){Rackham}, {Espinoza}, {Berdyugina}, {Korhonen}, {MacDonald}, {Montet}, {Morris}, {Oshagh}, {Shapiro}, {Unruh}, {Quintana}, {Zellem}, {Apai}, {Barclay}, {Barstow}, {Bruno}, {Carone}, {Casewell}, {Cegla}, {Criscuoli}, {Fischer}, {Fournier}, {Giampapa}, {Giles}, {Iyer}, {Kopp}, {Kostogryz}, {Krivova}, {Mallonn}, {McGruder}, {Molaverdikhani}, {Newton}, {Panja}, {Peacock}, {Reardon}, {Roettenbacher}, {Scandariato}, {Solanki}, {Stassun}, {Steiner}, {Stevenson}, {Tregloan-Reed}, {Valio}, {Wedemeyer}, {Welbanks}, {Yu}, {Alam}, {Davenport}, {Deming}, {Dong}, {Ducrot}, {Fisher}, {Gilbert}, {Kostov}, {L{\'o}pez-Morales}, {Line}, {Mo{\v{c}}nik}, {Mullally}, {Paudel}, {Ribas}, \& {Valenti}}]{Rackham2023}
{Rackham}, B.~V., {Espinoza}, N., {Berdyugina}, S.~V., {et~al.} 2023, RAS Techniques and Instruments, 2, 148, \dodoi{10.1093/rasti/rzad009}

\bibitem[{{Radica} {et~al.}(2025){Radica}, {Piaulet-Ghorayeb}, {Taylor}, {Coulombe}, {Benneke}, {Albert}, {Artigau}, {Cowan}, {Doyon}, {Lafreni{\`e}re}, {L'Heureux}, \& {Lim}}]{Radica2025}
{Radica}, M., {Piaulet-Ghorayeb}, C., {Taylor}, J., {et~al.} 2025, \apjl, 979, L5, \dodoi{10.3847/2041-8213/ada381}

\bibitem[{{Rathcke} {et~al.}(2025){Rathcke}, {Buchhave}, {de Wit}, {Rackham}, {August}, {Diamond-Lowe}, {Mendon{\c{C}}a}, {Bello-Arufe}, {L{\'o}pez-Morales}, {Kitzmann}, \& {Heng}}]{Rathcke2025}
{Rathcke}, A.~D., {Buchhave}, L.~A., {de Wit}, J., {et~al.} 2025, \apjl, 979, L19, \dodoi{10.3847/2041-8213/ada5c7}

\bibitem[{{Redfield} {et~al.}(2024){Redfield}, {Batalha}, {Benneke}, {Biller}, {Espinoza}, {France}, {Konopacky}, {Kreidberg}, {Rauscher}, \& {Sing}}]{Redfield2024}
{Redfield}, S., {Batalha}, N., {Benneke}, B., {et~al.} 2024, arXiv e-prints, arXiv:2404.02932, \dodoi{10.48550/arXiv.2404.02932}

\bibitem[{{Seager} \& {Sasselov}(1998)}]{Seager1998}
{Seager}, S., \& {Sasselov}, D.~D. 1998, \apjl, 502, L157, \dodoi{10.1086/311498}

\bibitem[{{Sing} {et~al.}(2011){Sing}, {Pont}, {Aigrain}, {Charbonneau}, {D{\'e}sert}, {Gibson}, {Gilliland}, {Hayek}, {Henry}, {Knutson}, {Lecavelier Des Etangs}, {Mazeh}, \& {Shporer}}]{Sing2011}
{Sing}, D.~K., {Pont}, F., {Aigrain}, S., {et~al.} 2011, \mnras, 416, 1443, \dodoi{10.1111/j.1365-2966.2011.19142.x}

\bibitem[{{Sudarsky} {et~al.}(2000){Sudarsky}, {Burrows}, \& {Pinto}}]{Sudarsky2000}
{Sudarsky}, D., {Burrows}, A., \& {Pinto}, P. 2000, \apj, 538, 885, \dodoi{10.1086/309160}

\bibitem[{{TRAPPIST-1 JWST Community Initiative} {et~al.}(2024){TRAPPIST-1 JWST Community Initiative}, {de Wit}, {Doyon}, {Rackham}, {Lim}, {Ducrot}, {Kreidberg}, {Benneke}, {Ribas}, {Berardo}, {Niraula}, {Iyer}, {Shapiro}, {Kostogryz}, {Witzke}, {Gillon}, {Agol}, {Meadows}, {Burgasser}, {Owen}, {Fortney}, {Selsis}, {Bello-Arufe}, {de Beurs}, {Bolmont}, {Cowan}, {Dong}, {Drake}, {Garcia}, {Greene}, {Haworth}, {Hu}, {Kane}, {Kervella}, {Koll}, {Krissansen-Totton}, {Lagage}, {Lichtenberg}, {Lustig-Yaeger}, {Lingam}, {Turbet}, {Seager}, {Barkaoui}, {Bell}, {Burdanov}, {Cadieux}, {Charnay}, {Cloutier}, {Cook}, {Correia}, {Dang}, {Daylan}, {Delrez}, {Edwards}, {Fauchez}, {Flagg}, {Fraschetti}, {Haqq-Misra}, {Huang}, {Iro}, {Jayawardhana}, {Jehin}, {Jin}, {Kite}, {Kitzmann}, {Kral}, {Lafreni{\`e}re}, {Libert}, {Liu}, {Mohanty}, {Morris}, {Murray}, {Piaulet}, {Pozuelos}, {Radica}, {Ranjan}, {Rathcke}, {Roy}, {Schwieterman}, {Turner}, {Triaud}, \& {Way}}]{TJCI2024}
{TRAPPIST-1 JWST Community Initiative}, {de Wit}, J., {Doyon}, R., {et~al.} 2024, Nature Astronomy, 8, 810, \dodoi{10.1038/s41550-024-02298-5}

\bibitem[{{Wakeford} {et~al.}(2019){Wakeford}, {Lewis}, {Fowler}, {Bruno}, {Wilson}, {Moran}, {Valenti}, {Batalha}, {Filippazzo}, {Bourrier}, {H{\"o}rst}, {Lederer}, \& {de Wit}}]{Wakeford2019}
{Wakeford}, H.~R., {Lewis}, N.~K., {Fowler}, J., {et~al.} 2019, \aj, 157, 11, \dodoi{10.3847/1538-3881/aaf04d}

\bibitem[{Yurchenko {et~al.}(2017)Yurchenko, Al-Refaie, \& Tennyson}]{Yurchenko2017}
Yurchenko, S.~N., Al-Refaie, A.~F., \& Tennyson, J. 2017, Astronomy \& Astrophysics, 605, A95, \dodoi{10.1051/0004-6361/201731015}

\bibitem[{{Zhang} {et~al.}(2018){Zhang}, {Zhou}, {Rackham}, \& {Apai}}]{Zhang2018}
{Zhang}, Z., {Zhou}, Y., {Rackham}, B.~V., \& {Apai}, D. 2018, \aj, 156, 178, \dodoi{10.3847/1538-3881/aade4f}

\bibitem[{Zieba {et~al.}(2023)Zieba, Kreidberg, Ducrot, Gillon, Morley, Schaefer, Tamburo, Koll, Lyu, Acu{\~{n}}a, Agol, Iyer, Hu, Lincowski, Meadows, Selsis, Bolmont, Mandell, \& Suissa}]{Zieba_2023}
Zieba, S., Kreidberg, L., Ducrot, E., {et~al.} 2023, Nature, \dodoi{10.1038/s41586-023-06232-z}

\end{thebibliography}
\bibliographystyle{aasjournal}

\end{document}